\documentclass[12pt]{article}
\usepackage{amsmath}
\usepackage{amssymb}
\usepackage{geometry}
\usepackage{epsfig}
\usepackage[usenames]{color}
\newcommand{\q}{\mathbf{q}}
\newcommand{\kk}{\mathbf{k}}
\newcommand{\dif}{\mathrm{d}}
\newcommand{\MSbar}{$\overline{\mathrm{MS}}\;$}
\newcommand{\absn}{\left(\frac{\bar\alpha_s}{N}\right)}
\usepackage{soul}

\def\pl#1#2#3{
        {\it Phys.\ Lett.\ }{\bf #1}, #2 (#3)}

\def\prl#1#2#3{
        {\it Phys.\ Rev.\ Lett.\ }{\bf #1}, #2 (#3)}

\def\pr#1#2#3{
        {\it Phys.\ Rev.\ }{\bf #1}, #2 (#3)}
\def\np#1#2#3{
        {\it Nucl.\ Phys.\ }{\bf #1}, #2 (#3)}

\def\jhep#1#2#3{
	{JHEP} {\bf #1}, #2 (#3)}

\begin{document}

\begin{flushright}
\mbox{
\begin{tabular}{l}
    IFUM-943-FT
\end{tabular}}
\end{flushright}
\vskip 1.5cm
\begin{center}
\Large
{\bf\sc High-energy resummation in direct photon production
 }

\vskip 0.7cm
\large
Giovanni Diana \\
\vskip 0.6cm
{\small Dipartimento di Fisica, Universit\`a di Milano and \\INFN, sezione di Milano, via Celoria 16, I-20133 Milan, Italy } \\
\vskip 0.6cm
\small\today
\end{center}
\thispagestyle{empty}
\vskip 0.7cm

\begin{abstract}
We present the computation of the direct photon production cross-section in perturbative QCD to all orders in the limit of high partonic center-of-mass energy. We show how the high-energy resummation can be performed consistently in the presence of a collinear singularity in the final state, we compare our results to the fixed order NLO cross-section in \MSbar scheme, and we provide predictions at NNLO and beyond.
\end{abstract}

\vspace{1cm}
\section{Introduction}
Perturbative QCD provides accurate theoretical predictions for hard processes at high-energy colliders. 
Logarithmic corrections to the lowest-order cross-sections can be systematically computed in the region of large hard scale $Q^2$, $\Lambda^2\ll Q^2\sim S$, by a renormalization group approach which leads to the factorization theorem of mass singularities~\cite{Altarelli:1977zs}. 
However, the TeV energy range opens up the two scale region $\Lambda^2\ll Q^2\ll S$, where the usual perturbative expansion receives large contributions characterized by logarithms of the ratio $x=Q^2/S$. 
In order to recover the accuracy of the perturbative results, logarithmically enhanced small-$x$ contributions to the hard cross-sections, associated to multiple gluon emission, must be resummed to all orders. 

Prompt photon production~\cite{old} is a relevant processes for the study of hard interactions in high-energy collisions. For example, it is the most important reducible background for the $H\rightarrow\gamma\gamma$ signal in the light Higgs scenario~\cite{Aad:2009wy}. A thorough understanding of this process in the small-$x$ limit is thus relevant to make predictions for the LHC.

Currently the direct photon cross-section is known up to $\mathcal{O}(\alpha\alpha_s^2)$~\cite{Gordon:1993qc} and Sudakov resummation effects have been computed up to NLL accuracy~\cite{Catani:1999hs}.
Prompt photon production is especially useful to probe the gluon parton density over a wide range of $x$~\cite{Aurenche:1988vi}, since the initial state gluon appears already at leading order,  and in particular the small-$x$ ($x\gtrsim 10^{-5}$) region. 
For this reason we expect that, in prompt photon production, high-energy resummation affects significantly the fixed-order results.

The general procedure for the small-$x$ leading-log (LL$x$) resummation of hard coefficient functions is well established in perturbative QCD within the framework of the $k_t$-factorization theorem~\cite{Catani:1990eg,Catani:1994sq}, and involves the computation of the leading amplitude of the process with off-shell incoming gluons. This technique has been used to obtain resummed cross-sections for heavy quarks photo- and hadro-production~\cite{Catani:1990eg,Ball:2007ra,Ball:2001pq,Camici:1997ta}, deep inelastic scattering~\cite{Catani:1994sq,Altarelli:2008aj}, Higgs production~\cite{Marzani:2008ih,Marzani:2008az,Hautmann:2002tu} and recently for the Drell-Yan process~\cite{Marzani:2008uh}.

Prompt photon cross-section contains two different contributions: the direct component, where the photon participates at leading order to the hard process, and a fragmentation component, which is needed to take account of the hadronic component of the photon. From a phenomenological point of view, at high-energy both terms are important~\cite{Gluck:1994iz}; in this work we will consider the direct contribution, leaving the fragmentation component to a future work.

All the processes for which small-$x$ resummation has been performed so far are free of collinear singularities in the final state since the corresponding cross-sections are totally inclusive; on  the contrary such a divergence does appear in direct photon production because the process is exclusive with respect to the final state photon, which from this point of view must be viewed as another hadronic state~\cite{Ellis:1991qj}. In this work we perform the high-energy resummation of the direct photon coefficient function consistently with the \MSbar scheme of subtraction of the final state singularity to all orders in perturbation theory.

\section{Prompt photon production}

\vspace{.5cm}
\subsection{Collinear factorization}
\vspace{.5cm}

The prompt photon process is characterized by a hard event involving the production of a single photon. Let us consider the hadronic process
\begin{equation}
H_1(P_1)+H_2(P_2)\rightarrow \gamma(q)+X.
\end{equation}
According to perturbative QCD, the direct and the fragmentation component of the inclusive cross-section at fixed transverse momentum $\q$ of the photon can be written as~\cite{Catani:1999hs}
\begin{eqnarray} 
&&\q^3 \frac{d\sigma_\gamma(x_\perp,\q^2)}{d\q}=\sum_{a,b}\int_{x_\perp}^1dx_1\; f_{a/H_1}(x_1,\mu^2_F)\int_{x_\perp/x_1}^1 dx_2\;f_{b/H_2}(x_2,\mu^2_F)\nonumber\times\\&&\times\int_0^1 dx\left\lbrace\delta\left(x-\frac{x_\perp}{x_1 x_2}\right)  \mathcal{C}^\gamma_{ab}(x,\alpha_s(\mu^2);\q^2,\mu_F^2,\mu_f^2)+\nonumber\right.\\
&&\left.+\sum_c\int dz\;z^2d_{c/\gamma}(z,\mu^2_f)\delta\left(x-\frac{x_{\perp}}{zx_1x_2}\right)\mathcal{C}^c_{ab}(x,\alpha_s(\mu^2);\q^2,\mu_F^2,\mu_f^2)\right\rbrace, \label{eq:fact}
\end{eqnarray}
where we have introduced the customary scaling variable:
\begin{equation}
x_\perp=\frac{4 \q^2}{S},  \qquad 0<x_\perp<1.
\end{equation}
in terms of the hadronic center-of-mass energy $S=(P_1+P_2)^2$. In the factorization formula eq.~(\ref{eq:fact}) we have used the short-distance cross-sections
\begin{eqnarray}
\mathcal{C}^{\gamma(c)}_{ab}&\equiv &\q^3\frac{d\hat\sigma_{ab\rightarrow\gamma(c)}(x,\alpha_s(\mu^2);\q^2,\mu_F^2,\mu_f^2)}{d\q},
\end{eqnarray}
where $a$, $b$ and $c$ are parton indices ($q$, $\bar q$, $g$) while $f_{i/H_j}(x_i,\mu^2_F)$ is the parton density at the factorization scale $\mu_F$. The fragmentation component is given in terms of a convolution with the fragmentation function $d_{c/\gamma}(z,\mu^2_f)$.

If we define the Mellin moments
\begin{eqnarray}
\sigma^\gamma(N)&=&\int_0^1 dx_\perp\; x_\perp^{N-1} \left(\q^3 \frac{d\sigma_\gamma(x_\perp,\q)}{d\q}\right),\nonumber\\
F_{i/H}(N,\mu_F^2)&=&\int_0^1 dx \;x^{N-1}\; x f_{i/H}(x,\mu_F^2),\nonumber\\
D_{c/\gamma}(N,\mu_f^2)&=&\int_0^1 dx\; x^{N-1}\; x^3 d_{c/\gamma}(x,\mu_f^2),\nonumber\\
\tilde{\mathcal{C}}^{\gamma(c)}_{ab}(N)&=&\int_0^1 dx\; x^{N-1}\, \mathcal{C}^{\gamma(c)}_{ab}(x)\nonumber
\end{eqnarray}
the collinear factorization theorem in $N$-space becomes
\begin{equation}
\sigma^\gamma(N)= \sum_{a,b} F_{a/H_1}(N) F_{b/H_2}(N) \left(\tilde{\mathcal{C}}^\gamma_{ab}(N)+\sum_c D_{c/\gamma} (N)\tilde{\mathcal{C}}^c_{ab}(N)\right).
\end{equation}

\vspace{.5cm}
\subsection{Leading-order coefficient functions and beyond}
\vspace{.5 cm}

At leading order the processes that contribute to the direct component of the prompt photon cross-section are
\begin{eqnarray} \label{eq:lo}
q\bar q\rightarrow \gamma g,\qquad q(\bar q) g\rightarrow \gamma q(\bar q).
\end{eqnarray}
The corresponding leading-order hard coefficient functions are given by~\cite{old,Ellis:1991qj}
\begin{eqnarray}
\mathcal{C}^{\gamma,LO}_{q\bar q}(x)&=&\frac{\q^3 d\hat\sigma_{q\bar q\rightarrow \gamma g}}{d\q}=\alpha\alpha_s Q^2_q\pi\frac{C_F}{C_A}\frac{x}{\sqrt{1-x}}(2-x), \label{eq:born1}\\
\mathcal{C}^{\gamma,LO}_{q(\bar q) g}(x)&=&\frac{\q^3 d\hat\sigma_{q(\bar q) g\rightarrow \gamma q(\bar q)}}{d\q}=\alpha\alpha_s Q^2_q\pi\frac{1}{2C_A}\frac{x}{\sqrt{1-x}}(1+\frac{x}{4})\label{eq:born2}
\end{eqnarray}
in terms of the partonic variable $x=4\q^2/\hat s$, where $\hat s$ is the partonic center-of-mass energy.
QCD corrections to the Born coefficient functions eqs.~(\ref{eq:born1}) and (\ref{eq:born2}) have been computed in ref.~\cite{ Gordon:1993qc} up NLO accuracy.

In the high-energy limit, the leading coefficient function is logarithmically enhanced by contributions of the form
 \begin{equation} \label{eq:nove}
\mathcal{C}_{qg}^\gamma(x)=\mathcal{C}^{\gamma,LO}_{qg}(x)+\alpha\alpha_s^2\sum_{k=0}^\infty c^{(k)}_{qg} (\alpha_s \log x)^k+\mathcal{O}\left(\alpha\alpha_s^3(\alpha_s\log x)^n\right)
 \end{equation}
which, in $N$-Mellin space, becomes a sum of poles at $N=0$ of increasing order:
\begin{equation} \label{eq:Npoles}
\tilde{\mathcal{C}}_{qg}^\gamma(N)=\tilde{\mathcal{C}}^{\gamma,LO}_{qg}(N)+\alpha\alpha_s\sum_{k=1}^\infty \tilde{c}^{(k)}_{qg} \left(\frac{\alpha_s}{N}\right)^k+\mathcal{O}\left(\alpha\alpha_s^2\left(\frac{\alpha_s}{N}\right)^n\right),\quad n\ge1.
 \end{equation}
As in the case of heavy quark production (HQ) and of the Drell-Yan processes (DY), the Born coefficients $\tilde{\mathcal{C}}^{\gamma,LO}_{ab}$ are regular as $N\rightarrow 0$: indeed, eqs.~(\ref{eq:born1},\ref{eq:born2}) vanish when $x$ approaches zero. 
The first singular term in eq.~(\ref{eq:Npoles}) is a simple pole in $N=0$ given by the NLO contribution to the perturbative series; in $x$-space, this  pole corresponds to a constant value, while the small-$x$ logarithms arise from the poles of increasingly higher order. 

\begin{figure}
\centering
\includegraphics[scale=.5]{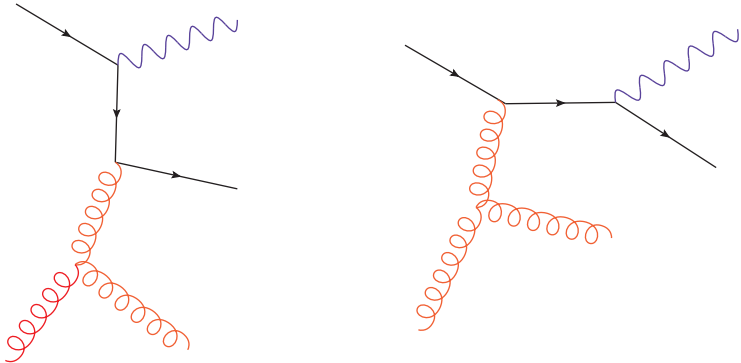}
\caption{In this picture are shown the dominant Feynman graphs in the high-energy limit. Both of the two  diagram determine the constant behaviour of the hard coefficient function $\mathcal{C}^\gamma_{q\bar g}$ at small-$x$. }\label{fig:2}
\end{figure}

\begin{figure}
\centering
\includegraphics[scale=.5]{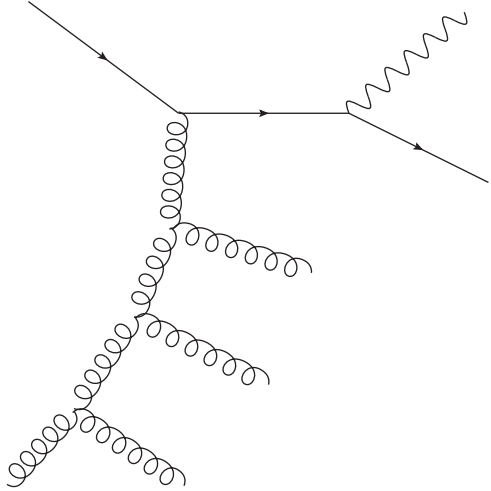}
\caption{Multiple gluon emission (BFKL ladder) from the initial state in direct photon production.}\label{fig:1}
\end{figure}

 The NLO contribution to the expansion given in eq.~(\ref{eq:nove}) $\alpha\alpha_s^2 c_{qg}^{(0)}$ has been computed for various processes in ref.~\cite{Ellis:1990hw}, in particular for direct photon production, by considering the Feynman diagrams in fig.~(\ref{fig:2}) where an extra gluon is radiated from the initial state. This is a general feature of the high-energy limit to all orders in perturbation thory: dominant contributions at high-energy (LL$x$) are given by the exchange of spin-1 particles in the $t$-channel therefore all the relevant Feynman diagrams in the small-$x$ limit are given by the BFKL ladders in fig.~(\ref{fig:1}).

From the NLO onwards, the direct photon cross-section acquires a final state divergence when the photon becomes collinear to the outgoing parton; this collinear divergence cannot be removed by adding the virtual corrections, rather it must be absorbed in the fragmentation component of eq.~(\ref{eq:fact}) as it happens for the initial state collinear divergences which are properly absorbed in the definition of the parton densities~\cite{Ellis:1991qj}.
In the next sections we will show how to remove this divergence in the \MSbar scheme consistently with the resummation procedure of the gluon ladder and we will give the analytic expression for the resummed hard coefficient function $\mathcal{C}^\gamma_{q\bar q}$ and $\mathcal{C}^\gamma_{q\bar g}$. 

\section{High-energy resummation}
In this section we compute the leading logarithmic corrections to the direct photon cross-section at high-energy.
The high-energy resummation of gluon ladders arises from the general formalism of the $k_t$-factorization theorem~\cite{Catani:1990eg,Catani:1994sq} and can be performed by following the computational procedure outlined in ref.~\cite{Marzani:2008az} which allows us to compute the coefficient of the dominant $\log x$ to all orders in perturbation theory.
This procedure requires the calculation of the Born cross-section  with off-shell (transverse) incoming gluons and off-shellness fixed in terms of their transverse momenta $\kk_i$ (impact factor).   

For a single off-shell gluon, we can parametrize the dependence on the virtuality $\kk^2$ through the dimensionless variable 
\begin{equation}
\xi={\kk^2\over Q^2},
\end{equation}
where $Q^2$ is the hard scale of the process which determines the argument of the running coupling $\alpha(Q^2)$. In direct photon production $Q^2$ is the magnitude of the transverse momentum of the photon $\q^2$. 
The LL$x$ resummation is performed by taking a Mellin transform of the off-shell cross-section
\begin{equation}
h(N,M)\equiv\int_0^\infty d\xi\; \xi^{M-1}\int_0^1 dx\;x^{N-1} \hat\sigma(x,\xi),
\end{equation}
and by identifying $M$ as the sum of leading singularities of the largest eigenvalue of the singlet anomalous dimension matrix~\cite{Jaroszewicz:1982gr} (BFKL anomalous dimension) 
\begin{eqnarray}
M&=&\gamma_s\left(\frac{\alpha_s}{N}\right)+\mathcal{O}\left(\frac{\alpha_s^2}{N}\right) \label{eq:gammas}\\
\gamma_s\left(\frac{\alpha_s}{N}\right)&=&\sum_{n=1}^{\infty} c_n \left(\frac{C_A\alpha_s}{\pi N}\right)^n,\qquad c_n=1,0,0,2\zeta(3).
\end{eqnarray}
This corresponds to the sum of the high energy contributions coming from all diagrams of fig.~(\ref{fig:1}).
Finally, the N$^k$LO coefficient of the maximum power of $\log x$ is given by expanding the impact factor in powers of $\alpha_s$.
\begin{figure}
\centering
\includegraphics[scale=.5]{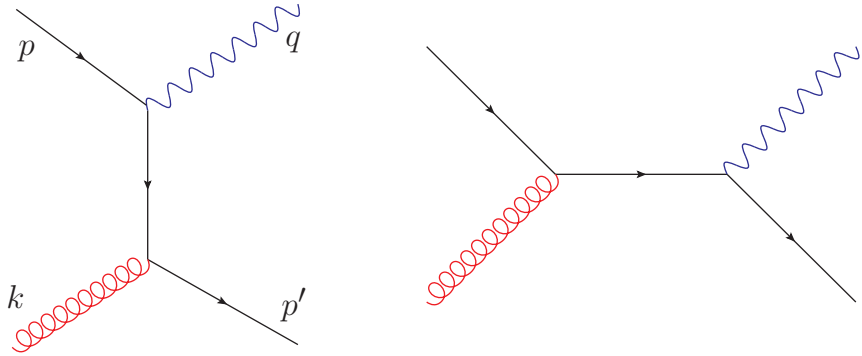}
\caption{Diagrams for direct photon production. } \label{fig:offshell}
\end{figure}

\vspace{.5cm}
\subsection{The off-shell cross-section}
\vspace{.5cm}

Let us consider the process 
\begin{equation}
g^\star (k)+q(p)\rightarrow \gamma(q)+q(p'),
\end{equation}
with an off-shell incoming gluon. 
We use a Sudakov parametrization for both incoming and outgoing momenta of the diagrams in fig.~\ref{fig:offshell}, thus we have
\begin{eqnarray}
p&=&z_2 p_2,\\
k&=&z_1 p_1+k_\perp,\\
q&=&x_1 z_1 p_1+(1-x_2)z_2 p_2+q_\perp,\\
p'&=&(1-x_1)z_1 p_1+x_2 z_2 p_2+k_\perp-q_\perp,
\end{eqnarray}
where
\begin{eqnarray}
k_\perp^2&=&-\vert\vec k_\perp\vert^2=-\kk^2\nonumber,\\
q_\perp^2&=&-\vert\vec q_\perp\vert^2=-\q^2\nonumber,
\end{eqnarray}
with light-like vectors $p_1$ and $p_2$ such that $p_1\cdot p_2=S/2$ where $S$ is the energy in the center-of-mass frame. 
The relevant scalar products are
\begin{eqnarray}
p\cdot k&=&s/2\nonumber,\\
p\cdot q&=&x_1 s/2\nonumber,\\
k\cdot q&=&(1-x_2)s/2-\vec k_\perp\cdot \vec q_\perp=\frac{\q^2}{2x_1} -\vec k_\perp\cdot \vec q_\perp\nonumber,
\end{eqnarray}
where we introduced the longitudinal energy $s=z_1 z_2 S$.
The $d$-dimensional phase space
\begin{eqnarray}
d\Phi^{(d)}&=&\frac{1}{(2\pi)^{d-2}}d^d q d^dp'\; \delta(p'^2)\delta(q^2)\delta^{(d)}(k+p-q-p')=\nonumber\\
     &=&\frac{1}{(2\pi)^{d-2}}d^d q\;  \delta((p+k-q)^2)\,\delta(q^2)
\end{eqnarray}
can be rewritten in terms of the Sudakov parameters since $d^{d} q=\frac{s}{2} d^{d-2}\mathbf{q}dx_1 dx_2$. We obtain
\begin{eqnarray}
d\Phi^{(d)}&=&\frac{1}{(2\pi)^2}\frac{s}{2} d^{d-2}\mathbf{q}dx_1 dx_2 \delta(-\mathbf{k}^2+s-x_1 s-(1-x_2)s+2\mathbf{k}\cdot\mathbf{q})\cdot\nonumber\\
&& \qquad\qquad\qquad\qquad\cdot\delta(-\mathbf{q}^2+x_1(1-x_2) s)\nonumber=\\
&=& \frac{(4 \pi)^{-\epsilon}\sqrt{\pi}}{(2\pi)^2}\frac{(\sin\theta)^{2\epsilon}}{\Gamma(1/2+\epsilon)}\frac{dx_1}{x_1}\; \q^{1+2\epsilon}d\mathbf{q}\;\delta(-\mathbf{k}^2+s(1-x_1)-\frac{\mathbf{q}^2}{x_1}+2\mathbf{k}\cdot\mathbf{q})
\end{eqnarray}
where $\theta$ is the angle between $\q$ and $\kk$ and we used the last $\delta$-function to perform the integration in $x_2$, now fixed to $x_2=1-\frac{\mathbf{q}^2}{x_1 s}$, which implies $x<4x_1$.  

Since we are interested in the differential cross-section  $\hat\sigma(x,\xi)=\q^3\frac{d\sigma}{d\q}$, our phase space in four dimension reduces to
\begin{equation}
\q^3 \frac{d \Phi^{(4)}}{d\q}= d \phi^{(4)}=\frac{1}{(2\pi)^2}\frac{dx_1}{2x_1}\;\q^4 d\theta\;\delta\left(-\mathbf{k}^2+s(1-x_1)-\frac{\mathbf{q}^2}{x_1}+2\mathbf{k}\cdot\mathbf{q}\right)\Theta(s-\kk^2),
\end{equation}
which in terms of the dimensionless partonic variables $x=4\q^2/s$ and $\xi=\kk^2/\q^2$ reads
\begin{eqnarray}
d\phi^{(4)}&=&\frac{1}{(2\pi)^2} \frac{xsdx_1}{8 x_1 }\;d\theta\;\delta\left(\frac{4}{x}(1-x_1)-\xi-\frac{1}{x_1}+2\sqrt{\xi}\cos\theta\right)\Theta(\frac{4}{x}-\xi)\Theta(4x_1-x)=\nonumber\\
&=&\frac{1}{(2\pi)^2} \frac{sxdx_1}{32x_1(1-x_1) }\;d\theta\;\delta\left(\frac{1}{x}-\frac{\xi+\frac{1}{x_1}-2\sqrt{\xi}\cos\theta}{4(1-x_1)}\right)\Theta(\frac{4}{x}-\xi)\Theta(4x_1-x).\nonumber
\end{eqnarray}

In the resummation procedure of refs.~\cite{Catani:1990eg,Catani:1994sq,Marzani:2008az} the computation of Feynman diagrams is performed 
by using the eikonal rule for the gluon polarization sum
\begin{equation}
\sum_\lambda \epsilon_\mu^\lambda(k)\epsilon_\nu^\lambda(k)=\frac{\kk_\mu\kk_\nu}{\kk^2},\qquad \kk_\mu\equiv(0,\kk,0),
\end{equation}
understood as the projector $\mathcal{P}$ over the high-energy singularities, analogously to the approach of refs.~\cite{Curci:1980uw}, which factorizes the gluon ladder from the Born coefficient.    
The channels $s$ and $t$ lead to the simple result for the amplitude in $d=4+2\epsilon$ dimensions 
\begin{eqnarray}\label{eq:ampl}
\mathcal{A}^{(d)}(x,x_1,\xi)&=&\overline{\sum}\mathcal{M}^2=\frac{4e^2 g_s^2}{2\cdot 2C_A}\left[\frac{(\q^2-s x_1)^2+s^2 x_1^2+\epsilon\q^4}{s x_1^3(s-\kk^2)}\right]\nonumber\\
&&\qquad=\frac{16e^2g_s^2}{2\cdot 2C_A}\left[\frac{1+\left(1-\frac{x}{4x_1}\right)^2+\frac{x^2}{16x_1^2}\epsilon}{x x_1\left(\frac{4}{x}-\xi\right)}\right]
\end{eqnarray}
averaged over color and helicity (of the quark) and summed over the final states.
As shown in eq.~(\ref{eq:Npoles}), the high-energy enhancement appears as a series of poles in $N=0$, therefore we are interested in the most singular term in the small-$N$ limit. 

Since the off-shell cross-section is well behaved at $N=0$, all the singular terms come from the substitution shown in eq.~(\ref{eq:gammas}), hence at this level we can reduce the computation to the $N=0$ moment of the impact factor in the $(N,M)$ space 
\begin{eqnarray}
h(0,M)&=&\frac{1}{(2\pi)^2}\frac{1}{2s}\int_0^{2\pi}\dif\theta\int_0^1\frac{s\;\dif x_1}{ 32x_1(1-x_1)}\int_0^{4x_1}\dif x\cdot\nonumber\\
&&\cdot\int_0^{4/x}\dif \xi\, \xi^{M-1} \mathcal{A}^{(4)}(x,x_1,\xi)\delta\left(\frac{1}{x}-\frac{\xi+\frac{1}{x_1}-2\sqrt{\xi}\cos\theta}{4(1-x_1)}\right)=\nonumber\\
%%&&=\frac{1}{(2\pi)^2}\frac{1}{2}\int_0^{2\pi}\dif\theta\int_0^1\frac{\dif x_1}{32x_1(1-x_1)}\int_{\frac{1}{4 x_1}}^{+\infty}\frac{\dif \rho}{\rho^2}\int_0^{4\rho}\dif \xi\, \xi^{M-1} \mathfrak{M}\delta\left(\rho-\frac{\xi+\frac{1}{x_1}-2\sqrt{\xi}\cos\theta}{4(1-x_1)}\right)=\nonumber\\
&=&\frac{1}{(2\pi)^2}\frac{1}{2}\int_0^{2\pi}\dif\theta\int_0^1\frac{\dif x_1}{32x_1(1-x_1)}\int_0^{\infty}\dif \xi\, \xi^{M-1}\cdot\nonumber\\
&&\cdot\int_{\max(\frac{1}{4x_1},\frac{\xi}{4})}^{+\infty}\frac{\dif \rho}{\rho^2}\mathcal{A}^{(4)}(1/\rho,x_1,\xi)\delta\left(\rho-\frac{\xi+\frac{1}{x_1}-2\sqrt{\xi}\cos\theta}{4(1-x_1)}\right),
\end{eqnarray}
where we introduced the variable $\rho=1/x$ and we exchanged the order of integration of $\xi$ and $\rho$.  By using the delta function we obtain:
\begin{eqnarray}
h(0,M)&=&\frac{1}{(2\pi)^2}\frac{1}{2}\int_0^{2\pi}\dif\theta\int_0^1\frac{\dif x_1}{32x_1(1-x_1)}\int_0^{\infty}\dif \xi\, \xi^{M-1}\cdot\nonumber\\
&&\cdot\frac{1}{\bar\rho^2}\mathcal{A}^{(4)}(1/\bar\rho,x_1,\xi)\Theta\left(\bar\rho-\max(\frac{1}{4x_1},\frac{\xi}{4})\right),\label{eq:bigtheta}
\end{eqnarray}
where we have defined
\begin{equation}
\bar\rho=\frac{\xi+\frac{1}{x_1}-2\sqrt{\xi}\cos\theta}{4(1-x_1)}.
\end{equation}

The argument of the Heaviside $\Theta$-function in eq.~(\ref{eq:bigtheta}) is always positive since
\begin{equation}
\bar\rho-\frac{1}{4x_1}=\frac{\xi x_1+1-2x_1\sqrt{\xi}\cos\theta-1+x_1}{4x_1(1-x_1)}=\frac{\xi-2\sqrt{\xi}\cos\theta+1}{4x_1(1-x_1)}>\frac{(\sqrt{\xi}-1)^2}{4x_1(1-x_1)}>0\nonumber
\end{equation}
when $1/x_1>\xi$, and
\begin{equation}
\bar\rho-\frac{\xi}{4}=\frac{\xi +\frac{1}{x_1}-2\sqrt{\xi}\cos\theta-\xi(1-x_1)}{4(1-x_1)}=\frac{\frac{1}{x_1}-2\sqrt{\xi}\cos\theta+\xi x_1}{4(1-x_1)}>\frac{(\frac{1}{\sqrt{x_1}}-\sqrt{\xi x_1})^2}{4(1-x_1)}>0\nonumber
\end{equation}
in the opposite case.
Therefore we have
\begin{eqnarray}  
h(0,M)&=&\frac{1}{(2\pi)^2}\frac{1}{2}\int_0^{\infty}\dif \xi\, \xi^{M-1} \int_0^{2\pi}\dif\theta\int_0^1\frac{\dif x_1}{32x_1(1-x_1)}\frac{1}{\bar\rho^2}\mathcal{A}^{(4)}(1/\bar\rho,x_1,\xi)\label{eq:unsub}.
\end{eqnarray}

The integration over the region $0<\theta<2\pi$, $0<x_1<1$   in eq.~(\ref{eq:unsub}) is always divergent when $\xi>1$, \emph{i.e.} $\vert\kk\vert>\vert \q\vert$; indeed the latter condition defines the kinematical region where the photon can be radiated collinearly to the quark in the final state. In the collinear limit, the amplitude in eq.~(\ref{eq:ampl}) is singular and the divergence is given by the fermionic propagator in the $s$-channel. In the collinear limit we have:
\begin{equation}
4\bar\rho-\xi=\frac{\xi x_1^2-2 \sqrt{\xi}x_1 \cos\theta +1}{x_1-x_1^2}=0,
\end{equation}
which happens when 
\begin{equation} \label{eq:singlim}
\left\lbrace\begin{array}{l}
       \theta=0\\  
	   x_1=\frac{1}{\sqrt\xi}
	   \end{array}\right. . 
\end{equation}
\vspace{.5cm}
\subsection{Subtraction of the collinear singularity in \MSbar scheme}
\vspace{.5cm}

In order to cancel the collinear divergence in the \MSbar scheme, 
first,
 we regularize the integrations of eq.~(\ref{eq:unsub}) by subtracting the collinear limit of the four dimensional amplitude before doing any integration, second we recover the pole in $\epsilon=0$ and the remaining finite parts by adding back the same quantity computed in $d$ dimensions.

We can do this by writing the impact factor in $4+2\epsilon$ dimensions as
\begin{eqnarray}
 h^{(d)}(x,\xi) &=&\int d\phi^{(d)}\, \mathcal{A}^{(d)}.
\end{eqnarray}
We then remove the singularity of the integrand by introducing a function $\mathcal{D}^{(d)}$ which has the same singular behaviour of the squared amplitute $ \mathcal{A}^{(d)}$ in the collinear limit eq.~(\ref{eq:singlim}). In four dimension we have 
\begin{equation}
\mathcal{D}^{(4)}=\frac{ e^2 g_s^2}{C_A} P_{q\gamma}(1/\sqrt{\xi}) \frac{\sqrt{\xi}-1}{(1-\sqrt{\xi} x_1)^2+\theta^2} \Theta(\xi-1),
\end{equation}
where
\begin{equation}
P(z)=\frac{1+(1-z)^2}{z}.
\end{equation}
By adding and subtracting the phase space integral of the function $\mathcal{D}^{(d)}$ to the $d$-dimensional impact factor we obtain
\begin{eqnarray}
h^{(d)}&=&\lim_{\epsilon\rightarrow 0} \left( \int d\phi^{(d)} \mathcal{A}^{(d)}  - \int d\phi^{(d)}\, \mathcal{D}^{(d)}\right)+\int d\phi^{(d)}\,\mathcal{D}^{(d)} + \mathcal{O}(\epsilon)=\nonumber\\
 &=&\int d\phi^{(4)} \left(  \mathcal{A}^{(4)}  -  \mathcal{D}^{(4)}\right)+ f_\mathcal{A}+\int d\phi^{(d)} \mathcal{D}^{(d)}+ \mathcal{O}(\epsilon) \label{eq:sub}
\end{eqnarray}
where the first integral is finite in four dimensions and the finite part $f_\mathcal{A}$ comes from the linear term in $\epsilon$ in the $d$-dimensional amplitude eq.~(\ref{eq:ampl}).

By using the $d$-dimensional phase space, the last term in eq.~(\ref{eq:sub}) is
\begin{equation}
 d\phi^{(d)} \mathcal{D}^{(d)}= \frac{\alpha \alpha_s}{2C_A}\frac{\sqrt{\pi}(4\pi)^{-\epsilon}}{\Gamma(1/2+\epsilon)}\left(\frac{\mu^2}{\q^2}\right)^{-\epsilon}\frac{1}{\xi} P_{q\gamma}(1/\sqrt{\xi}) \frac{\Theta(\xi-1)\delta\left(\frac{1}{x}-4x_1^2\right)}{ \left[(1-\sqrt{\xi} x_1)^2+\theta^2\right]}\; \theta^{2\epsilon}\,d x_1 d\theta,
\end{equation}
where the dimensional scale $\mu^2$ (introduced by dimensional regularization) from now on will be identified with $\q^2$. By using this result in eq.~(\ref{eq:sub}) and performing the Mellin integrations with $N=0$ we have
\begin{eqnarray}
h^{(d)}(0,\xi)&=&\frac{1}{(2\pi)^2}\frac{1}{2}\int_0^1\dif x_1\left(\int_{-\pi}^{\pi}\dif\theta \frac{1}{\bar\rho^2}\frac{\mathcal{A}^{(4)}(\bar\rho,x_1,\xi)}{32x_1(1-x_1)}-\int_{-\infty}^{\infty}\dif\theta \frac{1}{2\xi} \mathcal{D}^{(4)}\right)+f_{\mathcal{A}}+\nonumber\\
 &&+\frac{\alpha \alpha_s}{2C_A}\frac{\sqrt{\pi}(4\pi)^{-\epsilon}}{\Gamma(1/2+\epsilon)}\frac{1}{\xi}\int_0^1 dx_1\int_{-\infty}^{+\infty} d\theta   \frac{\theta^{2\epsilon}\,P_{q\gamma}(1/\sqrt{\xi})\Theta(\xi-1)}{ \left[(1-\sqrt{\xi} x_1)^2+\theta^2\right]}\label{eq:finalh} 
\end{eqnarray}
where we have extended the limits of the angular integration of the function $\mathcal{D}$ in order to simplify the results of the integration while the finite part $f_{\mathcal{A}}$ is
\begin{eqnarray}
%f_{\mathfrak{M}} &=&\frac{1}{(2\pi)^2}\frac{1}{2}\int \dif \xi\; \xi^{M-1}\int_0^1\frac{\dif x_1}{32x_1(1-x_1)}\left(\int_{-\pi}^{\pi}\dif\theta \frac{1}{\bar\rho^2}\frac{e^2g_s^2}{4 C_A}\frac{x^2 \epsilon}{x_1^3(4-\xi x)}\right)=\nonumber\\
f_{\mathcal{A}}&=&\lim_{\epsilon\rightarrow 0}\frac{\alpha\alpha_s\epsilon}{C_A}\int \dif \xi\; \xi^{M-1}\int_0^1 \dif x_1 \sqrt{\xi}\left(\int_{-\infty}^{+\infty}\dif\theta \frac{\xi^{-2} \; \theta^{2\epsilon}}{\left[(1-\sqrt{\xi} x_1)^2+\theta^2\right]}\right)=\nonumber\\
%&=&\frac{\alpha\alpha_s}{C_A}\int \dif \xi\; \xi^{M-1}\int \dif z \left(\int_{-\pi}^{\pi}\dif\theta \frac{\xi^{-2} \epsilon}{\left[z^2+\theta^2\right]^{1-\epsilon}}\right)=\nonumber\\
%&=&\frac{\pi\alpha\alpha_s}{C_A} \int_1^\infty \dif \xi\; \xi^{M-3}=
&=&\frac{\pi\alpha\alpha_s}{C_A} \frac{1}{2-M}. 
\end{eqnarray}

All the integrations in eq.~(\ref{eq:finalh}) can be performed in closed form, thus by subtracting the pole in $\epsilon=0$ with the usual combination $1/\epsilon+\gamma_E-\log4\pi$ we obtain\footnote{In ref.~\cite{Ellis:1990hw} the impact factor $h_q(a)$ was only computed in the
region $0<\xi<1$ where no collinear singularity is present. The result for $h_q$, given in eq.~(3.7) of ref.~\cite{Ellis:1990hw}, is seen to agree with our result recalling that in the region $0<\xi<1$ $h_q$ is related to $h$ eq.~(\ref{eq:hxispace}) by
\begin{equation}
\xi \frac{d h_q(\xi)}{d\xi}= h(\xi), \qquad (0<\xi<1)\nonumber.
\end{equation}
Note however that the
 expression for the cross-section $\sigma_{qg}(\q>p_T)$ given in eq.~(3.11) of
ref.~\cite{Ellis:1990hw} is
too large by a factor 2~\cite{ellpriv}.}
\begin{eqnarray}
h(0,\xi)&=&\frac{\pi\alpha\alpha_s}{C_A}\left\lbrace \Theta(\xi-1)\left(-\frac{1}{\xi}+\frac{8}{\sqrt{\xi}}(1-\log2)\right)+\right.\nonumber\\
&& \hspace{-1cm}\left.+~\mathrm{sign}(\xi-1)\left(3-\left(1+\frac{1}{2\xi}\right)\log(1-\xi)^2-\frac{1}{\sqrt{\xi}}\log\left(\frac{\sqrt\xi+1}{1-\sqrt\xi}\right)^2\right)\right\rbrace \label{eq:hxispace},
\end{eqnarray}
while the M-Mellin moments are
\begin{equation}
h(0,M)= \frac{\alpha\alpha_s\pi}{C_A}\left\lbrace \frac{(7-7M+2M^2) }{(M-1) (M-2)(2M-3)}\left( \pi  \cot (M \pi )+2  H_{M-2}+\frac{2}{M-1}\right)+\frac{1}{2-M}\right\rbrace,\label{eq:impfacM}
\end{equation}
where $H_{M-2}$ is the harmonic number of argument $M-2$.

\section{Results}

\vspace{.5cm}
\subsection{The resummed coefficient function}
\vspace{.5cm}

By expanding the impact factor obtained in the previous section around $M=0$, we obtain
\begin{eqnarray}
&&M h(0,M;\alpha_s)=\frac{\pi\alpha\alpha_s}{C_A}\left\lbrace\frac{7}{6 }+\frac{67}{36}M+\frac{385
   M^2}{216}+\left(\frac{2323}{1296}+\frac{7 \zeta
   (3)}{3}\right) M^3+\right.\nonumber\\
   &&{}+\left. \left(\frac{14233}{7776}+\frac{49 \zeta
   (3)}{18}\right) M^4
   +\left(\frac{87307}{46656}+\frac{331
   \zeta (3)}{108}+\frac{7 \zeta (5)}{3}\right)
   M^5+O\left(M^6\right)\right\rbrace.
\end{eqnarray}
Notice that in this formalism the collinear divergence from the initial state appears as a simple pole in $M=0$. The resummed coefficient function in the \MSbar factorization scheme is given by the relation  
\begin{equation}
\tilde{\mathcal{C}}^\gamma_{qg}(N,\alpha_s)=M h(N,M;\alpha_s) R(M)\big\vert_{M=\gamma_s}
\end{equation}
in terms of the impact factor eq.~(\ref{eq:impfacM}) and the function
\begin{equation}
R(M)=1+\frac{8}{3}\zeta_3 M^3-\frac{3}{4}\zeta_4M^4+\mathcal{O}(M^5)
\end{equation}
which takes into account finite parts coming from the \MSbar subtraction of initial state collinear singularities and where $M$ is identified as the BFKL anomalous dimension
\begin{equation}
\gamma_s\left(\frac{\bar\alpha_s}{N}\right)=\frac{\bar\alpha_s}{N}+2\zeta_3\left(\frac{\bar\alpha_s}{N}\right)^4 +2\zeta_5 \left(\frac{\bar\alpha_s}{N}\right)^6+\dots,\qquad \bar\alpha_s=\frac{\alpha_s C_A}{\pi}
\end{equation}

We have
\begin{eqnarray}
&&\tilde{\mathcal{C}}^\gamma_{qg}(N,\alpha_s)=\frac{\pi\alpha\alpha_s}{C_A}\left\lbrace\frac{7}{6}+\frac{67}{36}\absn+\frac{385
   }{216}\absn^2+\left(\frac{2323}{1296}+\frac{49 \zeta
   (3)}{9}\right) \absn^3+\nonumber\right.\\
   &&\left.\phantom{iuhiuhi}+\left(\frac{14233}{7776}-\frac{7
   \pi ^4}{720}+\frac{308 \zeta (3)}{27}\right)
   \absn^4+O\left(\absn^5\right)\right\rbrace.\label{eq:res1}
\end{eqnarray}
The NLO term in eq.~(\ref{eq:res1}) gives, in the $x$-space, the constant value 
$67/36\;\alpha\alpha_s^2$
which is in agreement with the fixed order calculation of refs.~\cite{Gordon:1993qc,Ellis:1990hw}.
By using the high-energy color charge relation between the hard coefficient functions
\begin{eqnarray}
\tilde{\mathcal{C}}^\gamma_{q\bar q(q)}(N,\alpha_s)=\frac{C_F}{C_A}\left(\tilde{\mathcal{C}}^\gamma_{qg}(N,\alpha_s)-\tilde{\mathcal{C}}^{\gamma,LO}_{qg}(0,\alpha_s)\right)
\end{eqnarray}
we can also obtain the LL$x$ contributions coming from the process $q\bar q\rightarrow \gamma g$
\begin{eqnarray}
&&\tilde{\mathcal{C}}^\gamma_{q\bar q(q)}(N,\alpha_s)=\alpha\frac{\alpha_s^2}{N}\frac{C_F}{C_A}\left\lbrace\frac{67}{36}+\frac{385
   }{216}\absn+\left(\frac{2323}{1296}+\frac{49 \zeta
   (3)}{9}\right) \absn^2+\nonumber\right.\\
   &&\left.\phantom{iuhiuhi}+\left(\frac{14233}{7776}-\frac{7
   \pi ^4}{720}+\frac{308 \zeta (3)}{27}\right)
   \absn^3+O\left(\absn^4\right)\right\rbrace.
\end{eqnarray}

\vspace{.5cm}
\subsection{Phenomenology}
\begin{figure}
\centering
\includegraphics[scale=.5]{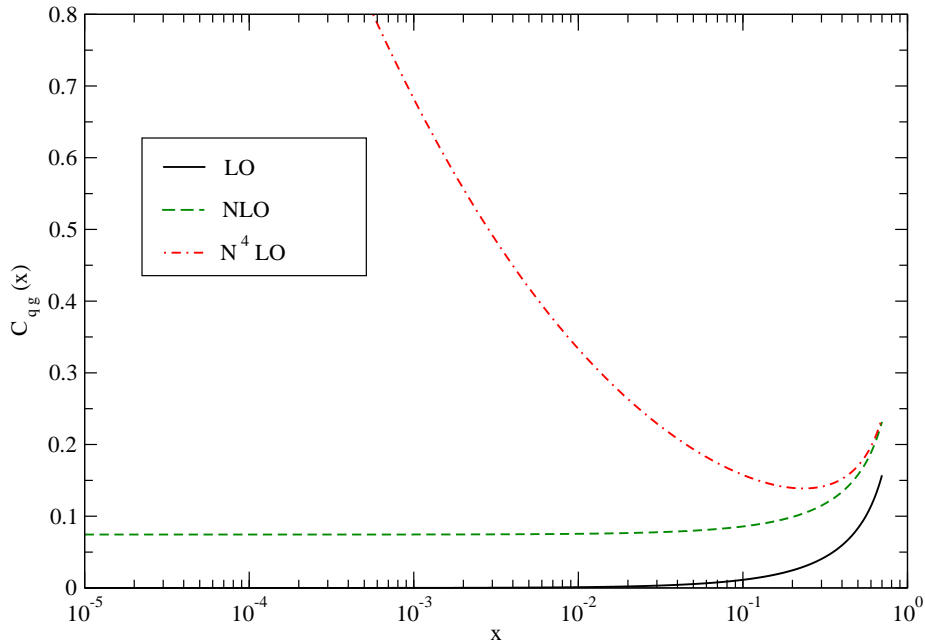}
\caption{The picture shows a comparison between the leading order direct photon coefficient function (black solid)  and the relative small-$x$ corrections up to NLO (green dashed) and N$^4$LO (red dashed). } \label{fig:3}
\end{figure}

\begin{figure}
\centering
\includegraphics[scale=.5]{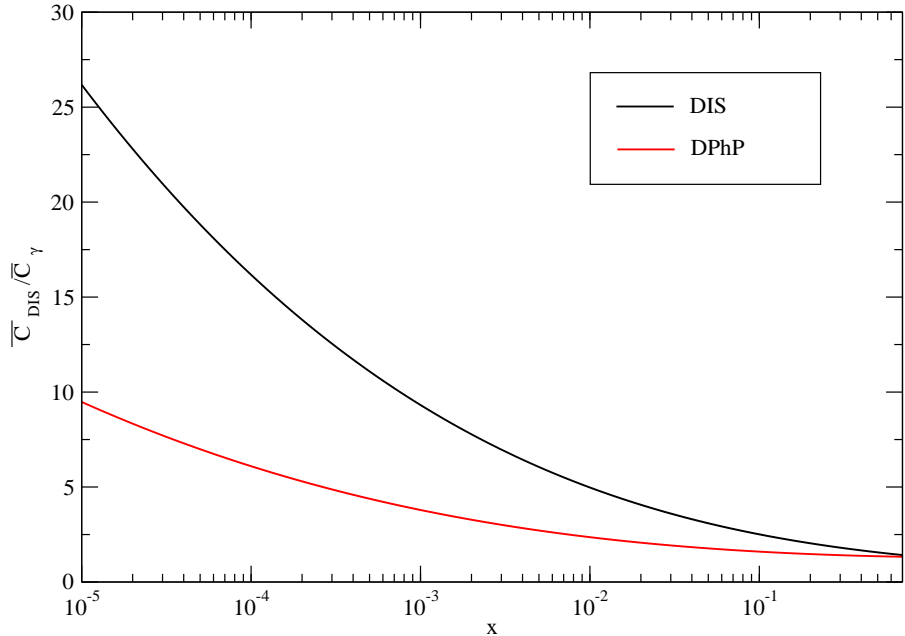}
\caption{Ratio between F$_2$ ($\bar{\mathcal{C}}_{DIS}$ and direct photon (DPhP) ($\bar{\mathcal{C}}_\gamma$) coefficient functions, both of them normalized to their LO values} \label{fig:4}
\end{figure}

\vspace{.5cm}
In fig.~\ref{fig:3} we compare in $x$-space the coefficient function $\mathcal{C}_{qg}(x,\alpha_s)$ at LO, NLO and N$^4$LO in the high-energy limit. The large contributions at small-$x$ spoil the perturbative expansion and must be resummed in order to recover accurate results. The resummation of these logarithmic terms in the hadronic cross-section can be performed to all orders~\cite{Altarelli:2008aj,Altarelli:2005ni} including running coupling.
A full phenomenological study could be performed by combining the resummed hard cross section computed here with the resummation of GLAP evolution equations, following the formalism of refs~\cite{Altarelli:2008aj,Altarelli:2005ni} (see also ref.~\cite{ciafaloni} for an alternative approach). However, we can get a feeling for the size of resummation effects by comparing the result in eq.~(\ref{eq:res1}) with the DIS coefficient function $C_2^g(N)$. As shown in fig.~\ref{fig:4} the ratio between $C_2^g$ and $\mathcal{C}_{qg}^\gamma$ (both of them normalized to the respective LO values) is of order 1, therefore we expect that, at low-$x$, resummation effects may be as important as those obtained in the DIS case.   

\section{Summary}
In this work we performed the computation of the hard coefficient function for direct photon production in the high center-of-mass energy limit (the small-$x$ limit) to all orders in perturbative QCD. We have shown how to resum the leading logarithms of the Bjorken variable $x$ which stem from the BFKL gluon ladders, and how the high-energy resummation technique can be joined with the \MSbar scheme of subtraction of a final state collinear singularities. 

High-energy resummation for both PDFs evolution and coefficient functions are well established tools which are needed to quantify the small-$x$ effects on perturbative results. A clear understanding of this high-energy effects will be crucial expecially for the LHC.

\section{Acknowledgements}
We thank Stefano Forte, who suggested this work, for useful discussions and critical reading of the manuscript, G.~Bozzi, A.~Vicini and S.~Marzani for several comments, F.~Caola for discussions and encouragement and A.~Sportiello for mathematical discussions. Special thanks are to R.~K.~Ellis and D.~A.~Ross for correspondence. 
This work was partly supported by the European
network HEPTOOLS under contract
MRTN-CT-2006-035505 and by a PRIN2006 grant (Italy).


\newpage

\begin{thebibliography}{99}


\bibitem{Altarelli:1977zs}
G.~Altarelli and G.~Parisi, \np{B126}{298}{1977}.

\bibitem{old}
H.~Fritzsch and P.~Minkowski, \pl{69B}{316}{1977};\newline
R.~Gastmans and T.~T.~Wu, {\it The Ubiquitous Photon} (Oxford, New York, 1990).

\bibitem{Aad:2009wy}
G.~Aad {\it et al.}, arXiv:0901.0512.

\bibitem{Gordon:1993qc} 
L.~E.~Gordon and W.~Vogelsang, \pr{D48}{3136-3159}{1993}.

\bibitem{Catani:1999hs}
S.~Catani, M.~L.~Mangano, P.~Nason, C.~Oleari and W.~Vogelsang, \jhep{03}{025}{1999}; \newline
P.~Bolzoni, S.~Forte and G.~Ridolfi, \np{B731}{85-108}{2005}.

\bibitem{Aurenche:1988vi}
P.~Aurenche, R.~Baier, M.~Fontannaz, J.~F.~Owens and M.~Werlen, \pr{D39}{3275}{89}.

\bibitem{Catani:1990eg}
S.~Catani, M.~Ciafaloni and F.~Hautmann, \np{B366}{135-188}{1991}.

\bibitem{Catani:1994sq}
S.~Catani and F.~Hautmann, \np{B427}{475-524}{1994}.

\bibitem{Ball:2007ra}
R.~D.~Ball, \np{B796}{137-183}{2008}.

\bibitem{Ball:2001pq}
R.~D.~Ball and R.~K.~Ellis, \jhep{05}{053}{2001}.

\bibitem{Camici:1997ta}
G.~Camici and M.~Ciafaloni, \np{B496}{305-336}{1997}.

\bibitem{Altarelli:2008aj}
G.~Altarelli, R.~D.~Ball and S.~Forte, \np{B799}{199-240}{2008}.

\bibitem{Marzani:2008ih}
S.~Marzani, R.~D.~Ball, V.~Del Duca, S.~Forte and A.~Vicini,
{\it Nucl. Phys. Proc. Suppl.} {\bf 186}, 98-101 (2009).

\bibitem{Marzani:2008az}
S.~Marzani, R.~D.~Ball, V.~Del Duca, S.~Forte and A.~Vicini,
\np{B800}{127-145}{2008}.

\bibitem{Hautmann:2002tu}
F.~Hautmann, \pl{B535}{159-162}{2002}.

\bibitem{Marzani:2008uh}
S.~Marzani and R.~D.~Ball, \np{B814}{246-264}{2009}.

\bibitem{Gluck:1994iz}
M.~Gluck, L.~E.~Gordon, E.~Reya and W.~Vogelsang,
\prl{73}{388-391}{1994}.

\bibitem{Ellis:1991qj}
R.~K.~Ellis, W.~J.~Stirling and B.~R.~Webber, "QCD and collider physics", {\it Camb. Monogr. Part. Phys. Cosmol.} {\bf 8}, 1-435, (1996).

\bibitem{Ellis:1990hw}
R.~K.~Ellis and D.~A.~Ross, \np{B345}{79-103}{1990}.

\bibitem{Jaroszewicz:1982gr}
T.~Jaroszewicz, \pl{B116}{291}{1982}.

\bibitem{Curci:1980uw}
G.~Curci, W.~Furmanski and R.~Petronzio, \np{B175}{27}{1980};\newline
R.~K.~Ellis, H.~Georgi, M.~Machacek, H.~D.~Politzer and G.~G.~Ross,
\np{B152}{285}{1979}.

\bibitem{ellpriv}
R.~K.~Ellis and D.~A.~Ross, Private communication.

\bibitem{Altarelli:2005ni}
G.~Altarelli, R.~D.~Ball and S.~Forte, \np{B742}{1-40}{2006}.

\bibitem{ciafaloni}
M.~Ciafaloni, D.~Colferai, G.~P.~Salam and A.~M.~Stasto,
\pr{D68}{114003}{2003}.


\end{thebibliography}
\end{document}